\documentclass[notitlepage,nofootinbib,showpacs,preprintnumbers,superscriptaddress,amsmath,amssymb,11pt]{revtex4-2}

\usepackage{graphicx}
\usepackage{dcolumn}
\usepackage{bm}
\usepackage{xcolor}
\usepackage[colorlinks=true,allcolors=blue]{hyperref}
\usepackage[normalem]{ulem}

\usepackage{lineno}

\begin{document}

\title{Beam Test Facilities for R\&D in Accelerator Science and Technologies}

\author{Christine Clarke}
\affiliation{SLAC National Accelerator Laboratory, Stanford University, Stanford, CA 94309 USA}

\author{Michael Downer}
\affiliation{University of Texas, Austin, TX 78705, USA}

\author{Eric Esarey}
\affiliation{Lawrence Berkeley National Laboratory, Berkeley, CA 94720, USA}

\author{Cameron Geddes}
\affiliation{Lawrence Berkeley National Laboratory, Berkeley, CA 94720, USA}

\author{Mark J. Hogan}
\affiliation{SLAC National Accelerator Laboratory, Stanford University, Stanford, CA 94309 USA}

\author{Georg Heinz Hoffstaetter}
\affiliation{Cornell University, Ithaca, NY 14853, USA}

\author{Chunguang Jing}
\affiliation{Euclid Techlabs LLC, Solon, OH 44139, USA}
\affiliation{Argonne National Laboratory, Argonne, IL 60439, USA}

\author{Steven M. Lund}
\affiliation{Michigan State University, East Lansing, MI 48824, USA}

\author{Sergei Nagaitsev}
\affiliation{Fermi National Accelerator Laboratory, Batavia, IL 60510, USA}
\affiliation{The University of Chicago, Chicago, IL 60637, USA}

\author{Mark Palmer}
\affiliation{Brookhaven National Laboratory, Upton, NY 11973, USA}

\author{Philippe Piot}
\affiliation{Northern Illinois University, DeKalb, IL 60115, USA}
\affiliation{Argonne National Laboratory, Argonne, IL 60439, USA}

\author{John Power}
\affiliation{Argonne National Laboratory, Argonne, IL 60439, USA}

\author{Carl Schroeder}
\affiliation{Lawrence Berkeley National Laboratory, Berkeley, CA 94720, USA}

\author{Donald Umstadter}
\affiliation{University of Nebraska-Lincoln, Lincoln, NE 68588, USA}

\author{Navid Vafaei-Najafabadi}
\affiliation{Stony Brook University, Stony Brook, NY 11794, USA}
\affiliation{Brookhaven National Laboratory, Upton, NY 11973, USA}

\author{Alexander Valishev}
\affiliation{Fermi National Accelerator Laboratory, Batavia, IL 60510, USA}

\author{Louise Willingale}
\affiliation{University of Michigan, Ann Arbor, MI 48109, USA}

\author{Vitaly Yakimenko}
\affiliation{SLAC National Accelerator Laboratory, Stanford University, Stanford, CA 94309 USA}

\begin{abstract}
This is the Snowmass Whitepaper on \emph{Beam Test Facilities for R\&D in Accelerator Science and Technologies} and it is submitted to two topical groups: AF1 and AF6.
\end{abstract}




\maketitle
\tableofcontents

\section{Executive Summary}
Demonstrating the viability of emerging accelerator science ultimately relies on experimental validation. A portfolio of beam test facilities at US National Laboratories and Universities, as well as international facilities in Europe and Asia, are used to perform research critical to advancing accelerator science and technology (S\&T).  These facilities have enabled the pioneering accelerator research necessary to develop the next generation of energy-frontier and intensity-frontier User Facilities. This report provides an overview of the current portfolio of beam test facilities outlining: the research mission, the recent achievements, and the upgrades required to keep the US competitive in light of the large investments in accelerator research around the world.

\emph{The mission of the US beam test facilities} is carried out by a broad community of accelerator scientists, drawn from Universities, Industry, and National Laboratories and has three core aims,

\begin{enumerate}
\item Provide the experimental testbeds where (1) exploratory accelerator research can be conducted (2) emerging accelerator research can be validated and (3) promising accelerator technology can be programmatically developed and undergo integration testing.
\item Develop accelerator S\&T needed to enable the next generation of HEP User Facilities for the energy- and intensity frontiers as well as other SC facilities such as light sources.
\item Educate and train future scientists and engineers.
\end{enumerate}

\emph{The capabilities of the US beam test facilities} considered in this report possess one or more of the following capabilities: ({\cal O}(100) MeV energy drive beams, {\cal O}(10) Petawatt drive lasers, {\cal O}(1) Gigawatt RF power sources, high-quality charged particle sources (e.g. low emittance electron beams), advanced beam manipulation systems (e.g. nonlinear integrable optics, optical stochastic cooling, emittance exchange) and capabilities to develop AI/ML for accelerator science.

\emph{Rapid progress was made by US beam test facilities} since the previous Snowmass Process. In 2015, the P5 strategic plan~\cite{p5report-2014} made multiple recommendations to the HEP General Accelerator R\&D (GARD) program and the beam test facilities were instrumental in addressing many of these. Most importantly, various roadmaps were successfully formulated by the GARD community in response to P5 to clearly articulate the long-term research needed to develop the next generation of energy-frontier and intensity-frontier machines. For example, the 2016 Advanced Acceleration Concepts (AAC) Roadmap identified a series of milestones, many of which were demonstrated at the US beam test facilities.  Milestones achieved include multi-GeV/m accelerating gradients, record-setting transformer ratios, the first demonstration of optical stochastic cooling, ion acceleration, etc.  In addition to HEP milestones, multiple spin offs from the beam test facilities have impacted mid-term applications, such as BES light-sources, with its role in the development of the RF photoinjector, demonstration of SASE FEL, etc. as well as near-term spinoffs impacting societal applications in the medical, security, and industrial fields with the development of compact accelerators.

\emph{Upgrades to US beam test facilities} will require significant investment by USDOE-HEP to ensure the US GARD program remains internationally competitive. In turn, upgraded beam test facilities are needed to enable DOE-HEP to build a cutting-edge program in the energy and intensity frontiers. For example, the European Strategy for Particle Physics has committed considerable funds to accelerator R\&D in Europe.  All USDOE GARD beam test facilities have near-term upgrades underway and have developed proposals for mid-term upgrades to continue progress on the previously developed roadmaps and support from DOE-HEP is needed to realize these plans. In the long-term, several options are under consideration ranging from a greenfield beam test facility to re-using the infrastructure of next-generation facilities such as ILC, CLIC, FNAL site linac, C3… etc with Advanced Acceleration technology. Once again, long-term plans require serious study and support. In the long term, the US community is exploring a possible greenfield beam test facility to enable a large-scale demo of the application to AAC and possibly support medium-energy research in elementary particle physics.

\section{Introduction}
In the United States, accelerator research and development (R\&D) is sponsored by the General Accelerator R\&D (GARD) Program of the Department of Energy (USDOE) Office of Science~\cite{usdoe}, National Science Foundation (NSF) programs~\cite{usnsf} and the Office of Accelerator Research \& Development (R\&D) and Production (ARDAP)~\cite{usdoe-ardap}. These programs support fundamental accelerator research and development of accelerator technology and methods to increase beam brightness, accelerating gradient, average power, etc., to dramatically improve the cost effectiveness and performance of accelerators~\cite{p5report-2014}. These programs are the key to creating accelerator User Facilities in the mid- and far-term for discovery science (particle physics, neutron sources, light sources, etc.) and to create near-term, low-cost, compact accelerator technology for future societal applications in the medical, security, and industrial fields. US beam test facilities are crucial components for advancing accelerator science and technology (S\&T). 

\section{Accelerator Beam Test Facilities Mission}
The mission of the beam test facilities are to provide the experimental test beds where fundamental accelerator research can be explored and advanced accelerator technology can be developed and undergo integration testing. While limited R\&D can be done in User Facilities dedicated to non-accelerator science (e.g. LHC), typically these facilities have a dedicated user base that cannot tolerate interruption.  Their limited beam availability is used for near-term accelerator technology to improve operations. These facilities are ill-suited for exploring and developing advanced accelerator science.  This is where beam test facilities, dedicated to accelerator science, play an essential role. In addition to the S\&T mission, beam test facilities are where a large number of future scientists and engineers needed by the accelerator community are educated and trained~\cite{eurostrat-report-2020,usdoe-aac-2016}. At the highest level, the mission of the facilities can be broken down as:
\begin{enumerate} 
\item Providing experimental test beds to carry out basic research in advanced accelerators and beam physics.
\item Developing the S\&T needed to enable the next generation of energy-frontier and intensity-frontier science facilities and societal accelerator applications.
\item Educating and training future scientists and engineers.
\end{enumerate}

%
%

\section{Overview of Existing Accelerator Beam Test Facilities}
The above mission is carried out at several beam test facilities located both domestically, at US National Laboratories and Universities and international facilities in Europe and Asia, are used to perform research critical to advancing accelerator science and technology (S\&T). In this section, a list of major beam test facilities that possess one or more of the following capabilities is presented: (O(100) MeV energy drive beams, O(10) Petawatt drive lasers, O(1) Gigawatt RF power sources, high-quality charged particle sources (low emittance electron beams and positron beams), advanced beam manipulation systems (e.g. nonlinear integrable optics, optical stochastic cooling, emittance exchange) and emerging capabilities to demonstrate AI/ML. Each facility description is accompanied by an overview of its research program. Note, this list is not exhaustive as it leaves out many university-scale beam test facilities due to space limitations. We note that university-based beam test facilities are vital to accelerator research since this is where many ideas are first developed before being tested on a larger scale.

\subsection{National Laboratories~\label{fac:labs}}

\subsubsection{The Accelerator Test Facility (ATF) at Brookhaven National Laboratory}
The Accelerator Test Facility (ATF) at Brookhaven National Laboratory serves the US DOE Accelerator Stewardship program and develops advanced acceleration methods for leptons and ions as well as high-energy photon sources. The ATF provides access to three classes of experimental facilities: a long-wave infrared high-power laser at 9.2 µm, a high-brightness linac-driven 75-MeV electron beam, and near-IR laser sources. This combination of capabilities enables programs on particle and photon source development, wakefield acceleration, inverse Compton scattering, and laser-driven plasma ion acceleration.
\begin{figure}[htbp]
\centering 
\includegraphics[width=0.95\textwidth]{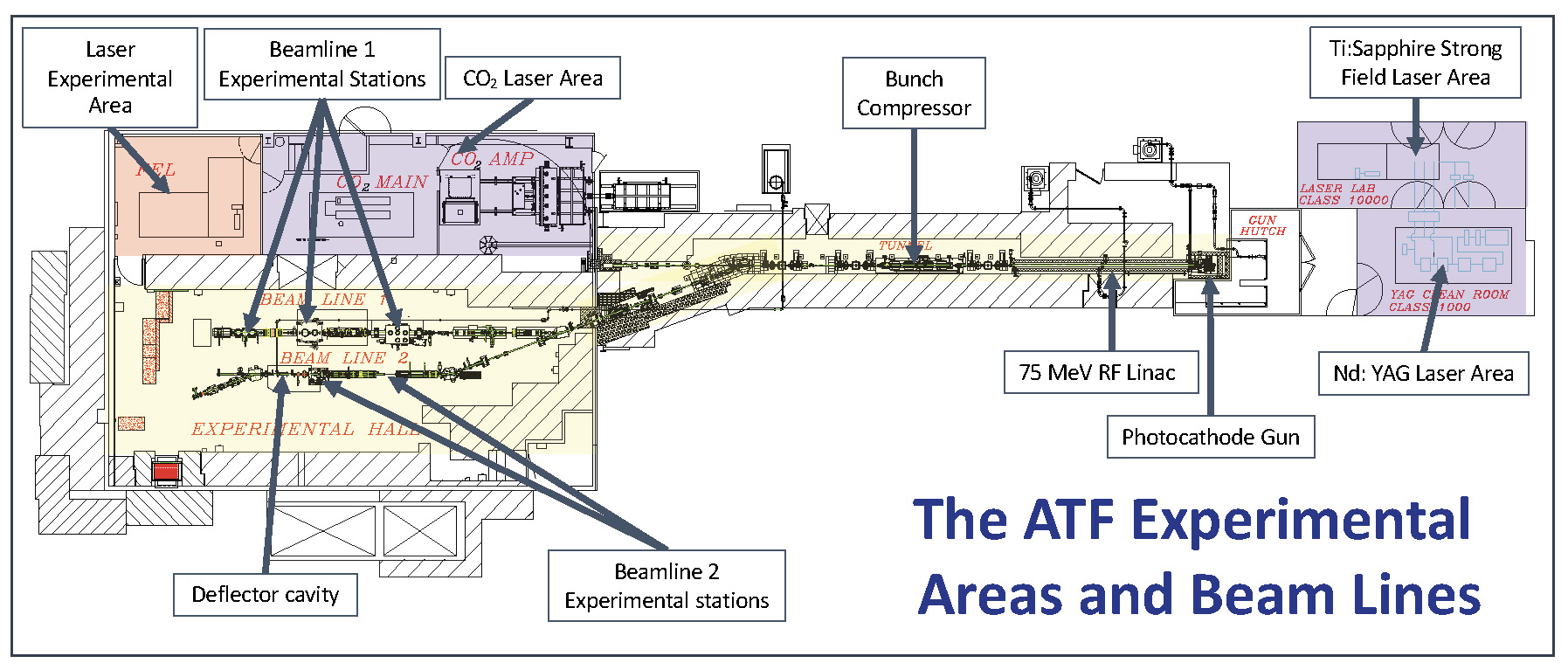}
\caption{\label{fig:atf} Overview of the ATF facility. User experiments can be configured to utilize the near-IR laser beams, long-wave IR beam, and electron beam singly or in any combination. }
\end{figure}

\subsubsection{The Argonne Wakefield Accelerator (AWA) at Argonne National Laboratory}
\begin{figure}[htbp]
\centering 
\includegraphics[width=.85\textwidth]{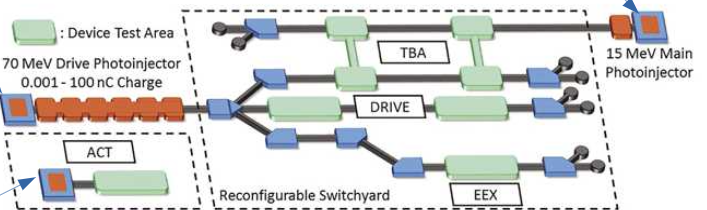}
\caption{\label{fig:awa} Overview of the AWA facility consists of a 70-MeV drive-beam photoinjector, a 15-MeV main-beam photoinjector, and the Argonne cathode test-stand (ACT).}
\end{figure}

The Argonne Wakefield Accelerator (AWA) Facility~\cite{awa-2022} is dedicated to the investigation of Structure Wakefield Acceleration (SFWA) including both collinear wakefield acceleration (CWA) and short-pulse two-beam acceleration (TBA). The AWA facility  (Fig.~\ref{fig:awa}) consists of three accelerators: the Drive Photoinjector (producing up to 70-MeV high-charge electrons), the Main Photoinjector (which forms 15-MeV electron bunches), and a few MeV cathode test-stand. AWA research focuses on underlying beam-dynamics challenges (e.g., production of bright- and high-charge-beams), beam manipulation (e.g., beam-current shaping) and beam control (e.g., mitigation of the BBU instability), and the development of advanced accelerating structures (e.g., dielectric waveguides, planar structures, metamaterials, rapid filling metallic structures, etc.) for efficient, high-peak-power RF generation and high-gradient acceleration. AWA has demonstrated unprecedented beam-transformer ratios ($>5$) in both SWFA and Plasma Wakefield Accelerator (PWFA) setups. Likewise, it has generated GW-scale peak-power RF-pulses and demonstrated staging in a TBA configuration in the pursuit of a linear collider. AWA develops these technologies for other applications (FELs and compact accelerators) and synergistically supports other advanced-accelerator concepts.

\subsubsection{The Berkeley Lab Laser Accelerator (BELLA) Center at Lawrence Berkeley National Laboratory}
\begin{figure}[htbp]
\centering 
\includegraphics[width=.85\textwidth]{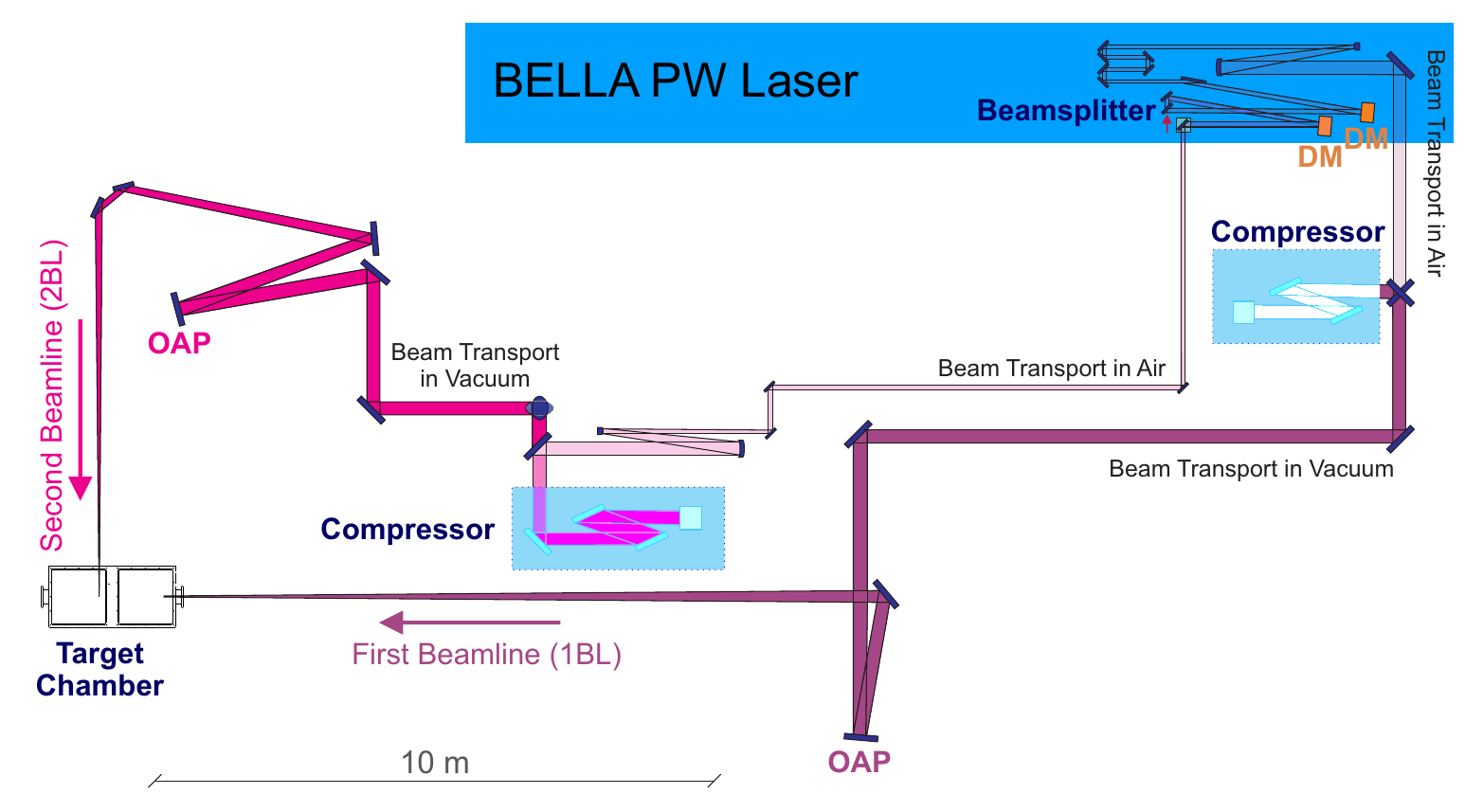}
\caption{\label{fig:bella} Schematic of the two beamline configuration on the BELLA PW laser.}
\end{figure}

The BELLA Center at LBNL has been performing research on laser-plasma accelerators (LPAs) for over two decades. The main research objectives are the development of LPA modules at the 10 GeV level and the staging (coupling) of LPA modules, which are two essential R\&D components for a future plasma-based linear collider. The current laser systems at the BELLA Center include the BELLA Petawatt (PW) laser and two independent 100 Terawatt (TW) laser systems. Upgrades are underway to the BELLA PW beamline to allow the delivery of two synchronized pulses, with independent compressors, on target, enabling staging, and a short focal length capability, enabling experiments at ultrahigh intensity (Fig.\ \ref{fig:bella}). The BELLA Center is also pursuing a new facility, kBELLA, consisting of a 1~kHz, few J, 30~fs, high average power laser for the demonstration of a high rep-rate, precision LPA and subsequent applications. The BELLA Center functions as a collaborative research center and is part of LaserNetUS~\cite{lasernetus}.

\subsubsection{The Fermilab Accelerator Science and Technology facility (FAST) at Fermi National Accelerator Laboratory}
\begin{figure}[htbp]
\centering 
\includegraphics[width=.9\textwidth]{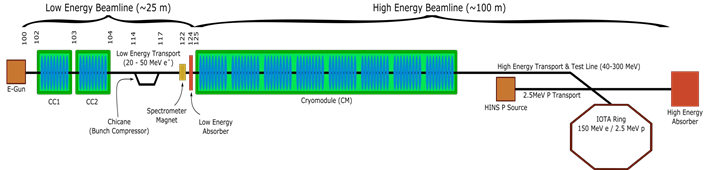}
\caption{\label{fig:FAST} Layout of the FAST facility.}
\end{figure}

The facility (Fig.\ \ref{fig:FAST}) consists of a storage ring, IOTA, capable of operating with both protons and electrons with beam momentum range 50-150~MeV/c and two injector linacs (electrons and protons). The IOTA research goals are mostly focused on the challenges, posed by future high-intensity machines, such as beam instabilities and losses. The IOTA storage ring is unique in its flexibility and performance. It has a circumference of 40 m and a relatively large aperture (50 mm). It is easily reconfigurable to accommodate the installation of 1-3 concurrent experiments. The focusing lattice was designed to have significant flexibility to enable a wide variety of studies. The IOTA proton injector produces beam at a fixed energy of 2.5 MeV (kinetic), while the energy output of the electron linac can be varied up to 300 MeV. Research with electron beams into optical stochastic cooling is underway at both IOTA (Fermilab) and CESR (Cornell).

More specifically, the IOTA research program aims to push the maximum beam intensity and brightness of future ring accelerators while minimizing the accelerator scale and cost. Along this direction, the key research areas are i) suppression of coherent beam instabilities by Landau damping; ii) mitigation of space-charge effects, and iii) beam cooling. The main candidate technologies being pursued at IOTA  are the {Nonlinear Integrable Optics}, {Electron Lenses}, and {Optical Stochastic Cooling}.

\subsubsection{The Facility for Advanced Accelerator Experimental Tests II (FACET-II) at SLAC National Accelerator Laboratory}
This facility is an upgrade of the FACET facility and operates as an Office of Science National User Facility. FACET-II will provide beams optimized for the next generation of PWFA experiments and will be the only facility in the world capable of providing 10 GeV electron beams in support of accelerator science R\&D. It allocates roughly half of the beam time towards investigating plasma wakefield acceleration in support of the R\&D roadmaps defined in the DOE Advanced Accelerator Development Strategy Report ~\cite{usdoe-aac-2016}. Primary elements of this program include demonstration of a 10 GeV plasma stage with preserved beam quality, development of ultra-high brightness beams from plasma-based injectors and developing high-intensity X-ray and Gamma-ray sources. The other half is dedicated to a diverse set of research programs enabled by high-energy high-intensity electron beams and their interaction with lasers, plasmas and solids. Novel diagnostics to characterize the extreme beams are being developed combining beam-physics, machine learning and artificial intelligence. Future upgrades to deliver 10 GeV high intensity positron beams and upgrades to the 10 TW experimental laser systems are under development to exploit the full scientific potential of the facility.

\begin{figure}[htbp]
\centering 
\includegraphics[width=1.00\textwidth]{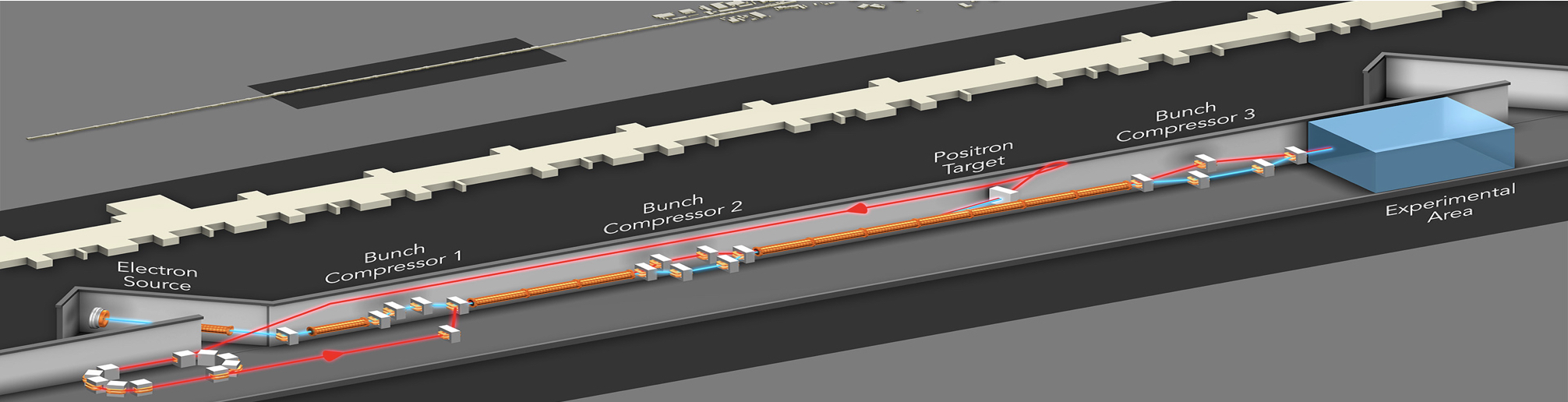}
\caption{\label{fig:facetII} Overview of the FACET-II facility. The existing facility includes the photoinjector (electron source), bunch compressors and experimental area (blue) and accelerators in between to achieve the nominal 10 GeV Energy. The facility has been designed to be compatible with upgrades to restore high-energy high-current positron bunches including a refurbished positron target and transfer line, a new compact damping ring and modified bunch compressors (red). The additional compressor upstream of the experimental area will allow simultaneous delivery of electrons and positrons to the experiments.}
\end{figure}

\subsection{Universities~\label{fac:uni}}

\subsubsection{The Cornell-BNL ERL Test Accelerator (CBETA)}
The Cornell-BNL ERL Test Accelerator (CBETA) consists of a 4-turn SRF Energy Recovery Linac with a permanent magnet Fixed Field Alternating-gradient (FFA) arc at Cornell University. The facility is equipped for studies of energy recovery, including phenomena such as the beam-breakup instability, beam timing and SRF phasing for simultaneous beam; for permanent magnet technology; and for FFA optics with large energy aperture. An additional focus is the saving of energy in accelerators.

\subsubsection{Extreme light laboratory, Lincoln, Nebraska}
The Extreme Light Laboratory (ELL) is home to the Diocles laser, a petawatt-class laser. It generates light with nearly a million-billion watts of power, or more than 100 times the combined output of all the world’s power plants, but compressed into the briefest of bursts. It can recreate some of the most extreme conditions in the universe. Experiments using Diocles enhance the basic understanding of physics and develop solutions for advanced manufacturing, national security and medicine. 
The lab houses two separate high-intensity femtosecond-duration lasers. One of them, the Diocles Laser, is petawatt class, and has been in operation for over 14 years. It is distinguished by exceptional laser beam quality, stability and parameter control. This enables research with high laser light intensity interacting with either plasmas, or GeV-energy electron beams from a coupled laser-driven wakefield accelerator. The interactions can be studied with synchronized pulses of laser light, electrons or tunable x-rays. A peak laser intensity of $6\times 10^{20}$ $6\times 10^{20}$ W/cm$^2$ was directly measured in an experimental demonstration of high-order multi-photon inverse Compton scattering.  
The laser and experimental research facility occupies 8,000-sq.-ft. three floors of the Behlen Lab building on the UNL City Campus. The laboratories meet stringent requirements on temperature, humidity, and vibration control, made possible by separate designated electrical, processed-chilled-water, and air-handling systems, and offers flexible experimental arrangements and diagnostics.  A total of six large 72”x48”x24” vacuum chambers are available, with three separate and independent beamlines: (1) 0.7-PW peak power at 0.1-Hz repetition rate, (2) 100-TW at 10-Hz and (3) 6-TW at 10-Hz. All use Ti:sapphire chirped-pulse amplification, and lase at 800-nm in 30-fs duration pulses.

\begin{figure}[htbp]
\centering 
\includegraphics[width=.65\textwidth]{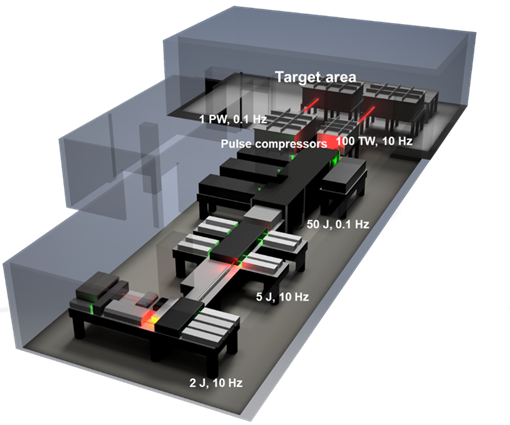}
\caption{\label{fig:dio} The Diocles laser at the Extreme Light Laboratory delivers peak intensity ranging from to $10^{18} - 10^{21}$ W/cm$^{2}$. }
\end{figure}

\subsubsection{Texas Petawatt Laser, University of Texas at Austin}
The Texas Petawatt Laser is a Nd:glass CPA laser using a unique architecture.  It combines amplification in an series of Optical Parametric Amplifiers and subsequent high energy amplification in a combination of silicate and phosphate glass to enable recompression of the pulses to under 140~fs, considerably shorter than most glass-based CPA lasers.  It can be fired into one of two target chambers, the first (TC1) focuses the beam tightly with an $f/3$ optic to intensity of well over 2 x $10^{21}$ W/cm$^{2}$.  An $f/1.1$ optic can be deployed in this chamber to yield intensity of over $10^{22}$ W/cm$^{2}$.  The beam can also be directed into a chamber with a long focal length, $f/40$ focusing.  This is optimal for gas target wakefield acceleration experiments.

\begin{figure}[htbp]
\centering 
\includegraphics[width=.65\textwidth]{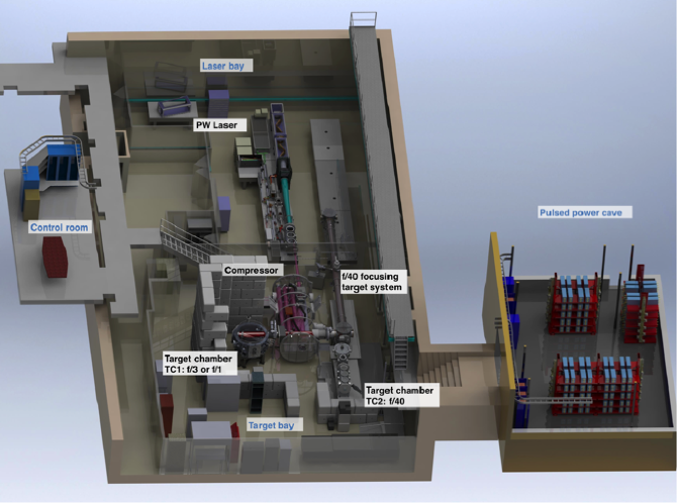}
\caption{\label{fig:dummy} Overview of the Texas Petawatt facility.}
\end{figure}

\subsubsection{ZEUS user facility, University of Michigan}
The National Science Foundation (NSF) Zettawatt Equivalent Ultrashort pulse laser System (ZEUS) will be a user facility located at the University of Michigan.
Due to become fully operational in late 2023, ZEUS will offer 30 weeks per year to users through a scientific merit based proposal system. There will be several modes of operation into three different target areas, each with different, flexible geometries to support a wide range of experiments. The full 3~PW power (75~J in 25~fs at 800~nm) will be delivered as a single pulse at a repetition rate of 1 shot per minute and should enable focused intensities exceeding $10^{22}~\rm{Wcm}^{-2}$. Alternatively, the power can be split into one beam with 2.5~PW and the with 500~TW. A signature configuration will use 2.5~PW to drive laser wakefield acceleration (LWFA) to multi-GeV electron energies and the 500~TW pulse will be tightly focused to counterpropagate to the electron beam. The apparent power of the focused laser pulse is boosted proportional to the square of the Lorentz factor (by more than a millionfold) from the Petawatt range to the Zettawatt range. Thus, the wakefield electrons see a Zettawatt Equivalent Ultrashort pulsed laser System – ZEUS.
Target Area 1 (TA1) will accommodate a very long focal length optic ($\sim f/60$) primarily designed to drive LWFA with up to 3~PW pulses. TA1 will also have a short focal length optic to create high-intensity conditions, that could be counter-propagating to the LWFA electron beam. Target Area 2 (TA2) will have a double plasma mirror capability, to create very high-contrast conditions that are desirable for high-harmonic generation or ion acceleration experiments using solid target interactions. Target Area 3 (TA3) will be operational at 500~TW at have a 5~Hz burst mode with $f/20$ or $f/40$ focusing, ideal for many LWFA or secondary betatron x-ray source studies and applications.

\begin{figure}[htbp]
\centering 
\includegraphics[width=.99\textwidth]{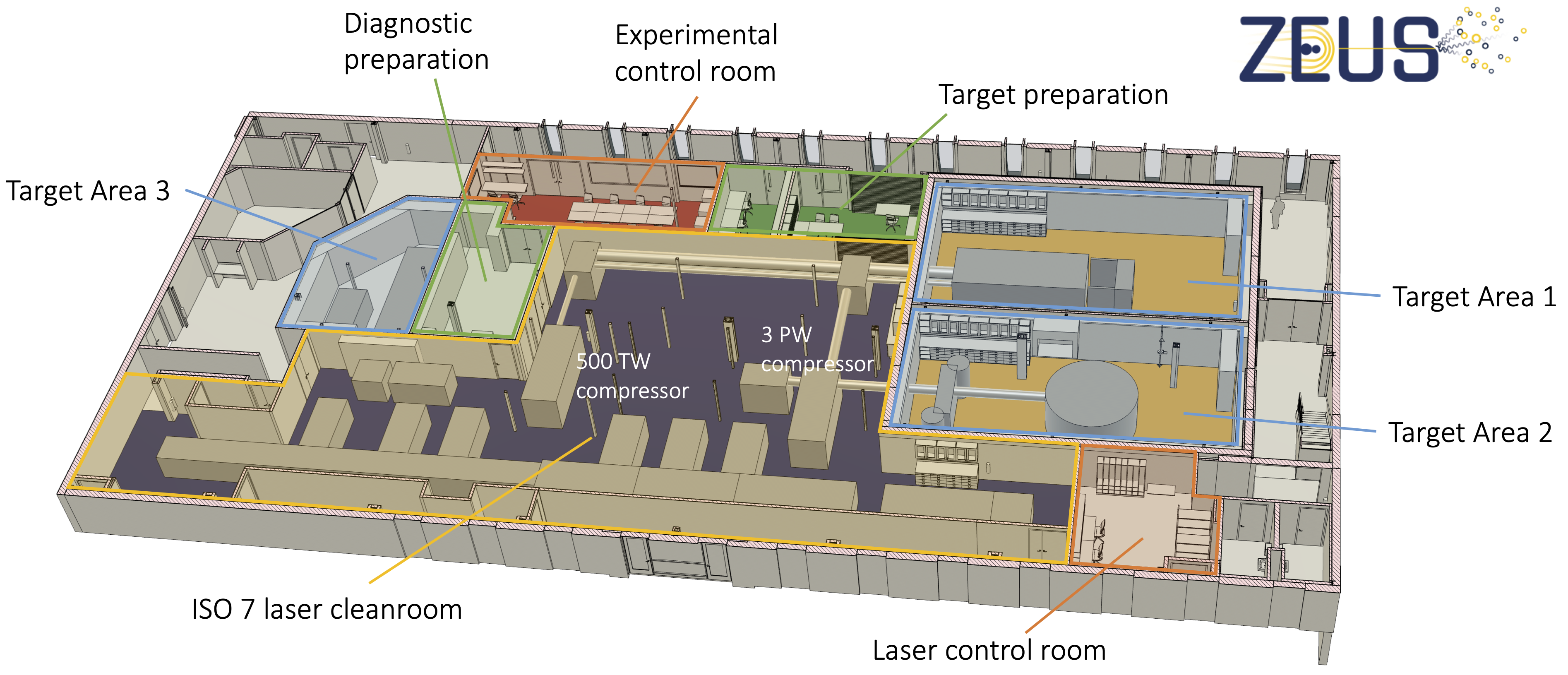}
\caption{\label{fig:zeus} Overview of the ZEUS facility.}
\end{figure}

%
%

\subsection{International Landscape~\label{fac:europe}}
Advancing accelerator science and technology (S\&T) is the subject of great international interest as the international community seeks to develop the next generation of energy-frontier and intensity-frontier user facilities.  This section provides a brief overview of beam test facilities around the world.

\subsubsection{European Facilities}
The European landscape currently maintains a large portfolio of Advanced and Novel Accelerator (ANA) beam test facilities for the development of technology which can be divided into two kinds: (i) beam-driven PWFA facilities and laser-driven laser wakefield acceleration (LWFA) facilities.  In addition, the recent European Strategy for Particle Physics ~\cite{adolphsen2022european} has committed considerable funds towards the commissioning of new ANA R\&D facilities. Muggli et al., (ref. JINST 076P 0122 Facility European) have published an article surveying this portfolio of current and future European ANA facilities.  Beam-driven PWFA facilities with the objective of generating high-energy (GeV or higher), high-quality beams, with established timeline and structured collaboration, typically beam  driven facilities, stemming from the accelerator community structuring include operating the facilities AWAKE ~\cite{Muggli_2017}, FLASHForward, and SPARC-LAB. Large scale PW laser facilities include a number of PW-class laser systems, although none are dedicated to R\&D towards accelerators for HEP. However, R\&D towards accelerators for HEP is part of the user program at many laser facilities, which significantly contribute to the development of high-energy accelerators.

In addition to the current ANA facilities, the European community has recently committed to the commissioning of a new large-scale accelerator project called EuPRAXIA.  In its first implementation phase, the EuPRAXIA consortium will construct a PWFA facility,
479 named EuPRAXIA@SPARC-LAB at the LNF-INFN~\cite{ferrario2018}. In its second phase, EuPRAXIA consortium will build a LWFA facility at a site to be chosen within the next 2 years between several options in Europe.  This major commitment by the European accelerator community will have a high impact on the future of accelerator technology.

\subsubsection{Asian Facilities~\label{fac:asian}}
The Asian accelerator landscape includes a variety of facilities to study the four ANA categories of LWFA, PWFA, structure wakefield acceleration (SWFA), and dielectric laser acceleration (DLA).  Kando et al., (ref. JINST 076P 0122 Facility European) have published an article surveying this portfolio of current and future Asian ANA facilities.  Most of the facilities are dedicated to LWFA including RIKEN SPring-8 in Japan and CoReLS in South Korea. The authors note that most of the Asian LWFA facilities possess high-peak-power (>10 TW) lasers with tens of femtosecond duration for laser wakefield acceleration.  There are also facilities at KEK and Tsinghua University in China that are capable of developing beam-driven PWFA and SWFA with novel structure development including dielectric lined circular waveguide (DLW), dielectric wall acceleration (DWA), and dielectric assisted accelerator (DAA).

%
%

\section{Research Thrusts}
Identifying the most promising accelerator research and development areas to be conducted at the ANA beam test facilities is an ongoing process and subject to continuous revision. For this reason, the US ANA beam test facilities are designed to provide access to a suite of complementary and diverse capabilities to serve a broad community of accelerator scientists drawn from Universities, Industry, and National Laboratories to carry out the national ANA mission. 

DOE and NSF work with US accelerator scientists to identify the most promising accelerator research thrusts and these findings have been published a number of documents. In 2016, HEP published the Advanced Accelerator Concepts (AAC) R\&D Roadmap~\cite{usdoe-aac-2016} for three AAC schemes, laser-driven plasma wakefield acceleration (LWFA), beam-driven PWFA, and structure wakefield acceleration (SWFA). The roadmaps describe a series of near-term, mid-term and long-term research milestones with the final goal of constructing a multi-TeV e+e– collider. Other important research thrusts include the Radiofrequency Accelerator  Roadmap~\cite{usdoe-rf-2017}, Accelerator and Beam Phyics Roadmap~\cite{abp-roadmap}, Accelerator Stewardship~\cite{usdoe-ardap}, Particle Sources, and Artificial Intelligence and Machine Learning~\cite{usdoe-aiml}.

\begin{table}[htbp]
\centering
\caption{\label{tab:i} Research Thrusts of the US National Laboratory ANA Beam Test Facilities.\label{tab:lab-thrust}}
\smallskip
\begin{tabular}{lccccc}
\hline
Research thrust & ATF & AWA & BELLA & FACET-II & FAST\\
\hline
{\bf Advanced Acceleration}  \\ 
\hline
LWFA & \checkmark (MeV) &   &\checkmark (GeV) & & \\
PWFA & \checkmark (MeV) & \checkmark (MeV)  & & \checkmark (GeV) & \\
SWFA & \checkmark (MeV) & \checkmark (MeV)  & & \checkmark (GeV) &  \\
IFEL & \checkmark &  & & & \\
staging & \checkmark & \checkmark & \checkmark & & \\
positron acceleration & & & & \checkmark & \\
\hline
{\bf Particle Source Development}  \\ 
\hline
plasma-based e- sources        &              & \checkmark &\checkmark  & \checkmark  &    \\
ion acceleration w/ lasers     & \checkmark   &            & \checkmark &    \\
Inverse Compton Scattering     & \checkmark   &            & \checkmark & \checkmark  &    \\
$\gamma$ ray via filamentation &              &            &            & \checkmark  &    \\
coherent X rays from plasmas   &              &\checkmark  &\checkmark         & \checkmark  &    \\
\hline
{\bf Beam Physics}  \\ 
\hline
phase-space cooling      &             & \checkmark   &           &            &  \checkmark  \\
single $e-$ \& crystalline beams &     &              &           &            &  \checkmark  \\
integrable nonlinear optics &          &              &           &            &  \checkmark  \\
extreme bunch compression    &          &              &           &\checkmark  &     \\
\hline
{\bf Diagnostics \& Beam Control}  \\ 
\hline
novel diagnostics            & \checkmark & \checkmark   &\checkmark  & \checkmark  &  \checkmark  \\
ML/AI: virtual diagnostics   &            & \checkmark   &\checkmark  & \checkmark  &  \checkmark  \\
ML/AI: improve efficiency    &            & \checkmark   &\checkmark  & \checkmark  &       \\
bunch-current shaping        & \checkmark & \checkmark   &\checkmark  & \checkmark  &       \\
phase-space/emit. exchange   & \checkmark &              &            &             &       \\
\hline
\end{tabular} 
\end{table}

In Table~\ref{tab:lab-thrust}, we list the main thrusts of the US ANA program as carried out by the ANA facilities at the national laboratories: advanced acceleration concepts, particle source development, beam physics, and diagnostics and beam control. Note that this list is not exclusive of all accelerator S\&T conducted at the US ANA facilities.

%
%

\section{Past Accomplishments}
ANA beam test facilities have been instrumental in advancing accelerator S\&T throughout their history. In this section, we present short summaries of these advances in two subsections (i) progress made since the previous Snowmass process and (ii) spinoffs from the DOE GARD beam test facilities.

\subsubsection{Accelerator S\&T: Progress Made since the Previous Snowmass Process}
The primary goal of the US DOE GARD program is to carry out accelerator-based particle physics research in support of the priorities  defined in the P5 strategic plan~\cite{p5report-2014}.  At the highest level, GARD supports three different lines of accelerators: the Intensity-Frontier with intense protons, the Energy-Frontier with a hadron collider and the Energy-Frontier with an e+e- linear collider. To accomplish this, a set of recommendations were made to the HEP GARD program by the HEPAP Accelerator R\&D Subpanel Report published in April 2015~\cite{hepap-subpanel-report-2015}.  Specifically, the subpanel published 15 recommendations for the GARD program and beam test facilities were instrumental in addressing 7 of these recommendations.

An comprehensive update on all of the GARD subpanel recommendations was presented at the 2019 November HEPAP meeting~\cite{hepap-meeting-2019}.  In this subsection, we present selected S\&T highlights of progress made on the subpanel recommendations in which the GARD beam test facilities were crucial. Examples of milestones accomplished covered here will include progress in both the intensity and energy frontiers. Intensity-Frontier accelerator science saw the first demonstration of Optical Stochastic Cooling~\cite{valishev-2021}. Energy-Frontier linear collider accelerator science delivered multi-GeV/m accelerating gradients and multi-GeV energy gain in plasma accelerators using beam-drivers at FACET~\cite{litos-2014} and laser-drivers at BELLA~\cite{gonsalves-2019}, multi-GeV/m positron acceleration at FACET~\cite{corde-2015}, staging of two plasma accelerator modules at BELLA~\cite{steinke-2016}, demonstration of controlled injection in a plasma wakefield accelerator~\cite{deng-2019}, record setting transformer ratios at AWA in both structures~\cite{gao-2018} and plasmas~\cite{roussel-2020}, pioneering work at ATF on shock wave monoenergetic ion acceleration~\cite{palmer-2011}, nonlinear effects in inverse Compton scattering (red shift, higher harmonics)~\cite{babzien-2006}, high-gain high-harmonic generation FEL~\cite{yu-2000}, fundamental research of electron acceleration in dielectric waveguides (e.g., using 3D woodpile structures~\cite{hoang-2018}), fundamental research on nonlinear Thomson scattering  \cite{fruhling-2021}, and novel methods for controlled optical injection of electrons into wakefields \cite{golovin2018electron}.  

\subsubsection{Community-Developed Roadmaps}
The development of community-driven research roadmaps was major organizational success for the GARD program.  Roadmaps have been published for all GARD thrusts except the Accelerator and Beam Physics Roadmap~\cite{nagaitsev2021accelerator} thrust area which was delayed by COVID but is expected to be published shortly.  These roadmaps enable pressing challenges to be more easily identified and addressed to move the field forward and insured a list of prioritized milestones that were aligned to the most compelling HEP science. The GARD beam test facilities play crucial roles in the following roadmaps. (i) In 2016, US DOE published the Advanced Accelerator Roadmap~\cite{usdoe-aac-2016} that laid out a series of research plans and goals that would provide the foundation for a technical design report of a multi-TeV linear collider. Significant progress along the AAC roadmap has been made at the GARD beam test facilities over the last 5 years and they remain critical for pursuing the continued advancements.

\subsubsection{Spinoffs from DOE GARD Beam Test Facilities}
A secondary, yet vital goal, is to apply this long-term, general R\&D carried out at the ANA facilities to generate spinoff technologies to benefit other applications in science, medicine and industry. Not only does this help validate the accelerator S\&T underway but also to benefit other accelerator applications (e.g. light sources, and neutron sources, etc.). Note that these spinoffs often come many years down the road, sometimes decades later. Sometimes they come from Universities, sometimes from National Laboratories, sometimes from Industry, or a combination of two or three stakeholders. A partial list of some of these spinoff technologies is given in Table \ref{tab:spinoff}.

%
%
\begin{table}[htbp]
\centering
\caption{R\&D spinoffs developed and/or demonstrated, at least in part, at US-ANA facilities. \label{tab:spinoff}}
\smallskip
\begin{tabular}{p{0.85\linewidth}}
\hline
Photoinjector R\&D that enabled high brightness sources for X-ray Free Electron Lasers (XFELs) \\
\hline
SASE/FEL R\&D demonstrating that the self-amplified spontaneous emission (SASE) instability would work at progressively shorter wavelengths from optical to X-rays \\
\hline
Inverse Free Electron Laser (IFEL) techniques that are basis for the laser heaters used to mitigate instabilities in XFELs\\
\hline
LWFA research that gave rise to betatron X-ray probes at the LCLS Matter in Extreme Conditions (MEC) Instrument\\
\hline
Laser scattering from plasma accelerated beams to generate X-ray and gamma ray probes \\
\hline
Wakefield studies that then were applied to create de-chirpers for XFELs and passive deflectors for UED facilities\\
\hline
Bunch compression studies that informed compressor design for higher peak current and shorter gain length at XFELS \\
\hline
Thomson scattering to measure electron beam emittance and bunch length  \\
\hline
Paricle (electron, proton and ion) beams from laser plasma accelerators for probing matter  \\
\hline
\end{tabular} 
\end{table}

%
%

\section{Capabilities}
The five US ANA beam test facilities, supported by the DOE Office of Science, HEP General Accelerator R\&D (GARD), provide access to a suite of complementary and diverse capabilities, as summarized in Table~\ref{tab:labs-capabilities}. Similarly, the capabilities of the university-based facilities are summarized in Table~\ref{tab:univ-capabilities}. Capabilities with high-quality beam parameters and diagnostic tools are essential to support research into new beam generation, acceleration, and transport techniques with the potential to mitigate technical risks of upcoming projects and to reduce the cost of future accelerator facilities.

%
%
\begin{table}[htbp]
\centering
\caption{Capabilities: National Laboratory ANA Beam Test Facilities. The ``P" label indicates a planned capability. \label{tab:labs-capabilities}}
\smallskip
\begin{tabular}{lccccc}
\hline
Capabilities & ATF & AWA & BELLA & FACET-II & FAST\\
\hline
{\bf operation model}  \\ 
\hline
National user facility   & \checkmark  &            &            &  \checkmark & \\
Accelerator stewardship  & \checkmark  &            &            & \checkmark  & \\
Collaboration            &             & \checkmark & \checkmark &             & \checkmark  \\
\hline
{\bf Beams \& Accelerators}  \\ 
\hline
$\sim 100$-MeV $e^-$       & \checkmark & \checkmark &\checkmark &            & \checkmark \\
$\sim 10$-GeV $e^-$        &            &            &  P        & \checkmark &            \\
$\sim 10$-GeV $e^+$        &            &            &           & P          &            \\
high-charge ($\sim 100$~nC) $e^-$  bunches  & & \checkmark &     &            &            \\
proton beams               &            &            &  P        &            & P  \\
NCRF S-band and X-band     & \checkmark &            &           &            &    \\
NCRF L-band and X-band     &            & \checkmark &           & \checkmark &    \\
SCRF L-band and X-band     &            &            &           &            & \checkmark \\
storage ring               &            &            &           &            & \checkmark \\
\hline
{\bf Lasers}  \\ 
\hline
TW-class 800-nm laser (Ti:Sapphire)    & \checkmark & \checkmark & \checkmark & \checkmark &   \\
PW-class 800-nm laser (Ti:Sapphire)    &            &            & \checkmark &            &    \\
TW-class 10~\textmu{m} laser (CO$_2$)  & \checkmark  &            &            &           &    \\
\hline
{\bf Plasmas}  \\ 
\hline
plasma capilaries (length [cm])   & \checkmark (2) & \checkmark (2)   &\checkmark (20)  &     &  \\
gas jets                     & \checkmark &        & \checkmark  & \checkmark  &      \\
heat-pipe oven               & \checkmark &        & \checkmark  & \checkmark  &       \\
hollow channel               &          & \checkmark   &         & \checkmark  &       \\
\hline
\end{tabular} 
\end{table}
%
%

%
%
\begin{table}[htbp]
\centering
\caption{Capabilities: University Beam Test Facilities. The ``P" label indicates a planned capability. \label{tab:univ-capabilities}}
\smallskip
\begin{tabular}{lccc}
\hline
Capabilities & ELL & Texas & ZEUS \\
\hline
{\bf Beams \& Accelerators}  \\ 
\hline
$\sim 100$-MeV $e^-$       & \checkmark & \checkmark &  P\\
$\sim 10$-GeV $e^-$        &  &  &  P\\
proton beams               & \checkmark & \checkmark &  P \\
\hline
{\bf Lasers}  \\ 
\hline
TW-class 800-nm laser (Ti:Sapphire)    & \checkmark &  &  P \\
PW-class 800-nm laser (Ti:Sapphire)    & \checkmark &  &  P \\
PW-class 1057-nm laser (Nd:glass) &  & \checkmark & \\
\hline
{\bf Plasmas}  \\ 
\hline
plasma capilaries (length [cm])   &  &  &  \\
gas jets                     & \checkmark & \checkmark &  P \\
heat-pipe oven               & &  &  \\
hollow channel               & & &  \\
\hline
\end{tabular} 
\end{table}
%
%
\section{ANA Beam Test Facility upgrades}

In order to continue progress in accelerator science and technology, for instance progress along the AAC Roadmap, major upgrades to the facilities are required.

\subsection{ATF}
The Accelerator Test Facility (ATF) at Brookhaven National Laboratory. The upgrade plan for ATF facilities reflects the priority set by the science planning workshop~\cite{atf-2019} based on community input. At the core of this plan is upgrading of the LWIR laser with the goal of delivering reliable CO$_2$ laser pulses at high repetition rate with a power of $>25$~TW and a sub-ps pulse length to the users in the long-term. Such intense LWIR sources (reaching $a_0$ as high as 8 with $F_\# \sim 1$ focusing) fill a unique niche in advanced acceleration research for generation of electron and ion beams~\cite{pogorelsky-2016,schroeder-2015,fiuza-2015}. While the facility is currently capable of producing a 5~TW CO$_2$ pulse, only about half of that is delivered to users due to limitations of the air transport system. An upgrade to a vacuum transport system as well as the development of nonlinear pulse compression (NLPC)~\cite{polyanskiy-2021} technique is expected to allow the delivery of >10 TW pulses over the next several years with pulse lengths as short as hundreds of femtoseconds. The facility is also pursuing an active R\&D program on optical pumping of CO$_2$ molecules as well as other technologies that make efficient use of the CO$_2$ gain profile and enable the delivery of high-repetition rate, high-average power, reliable LWIR pulses. The ATF plans also envision several thrusts for upgrading the linac-driven e-beam, including beam energy and pulse compression. Together with the expanded near IR capabilities, which encompasses the integration and the delivery of a compressed ($\le 100$~fs), terawatt-class Ti:Sapphire laser pulse to the IP, simultaneous experiments using the LWIR laser, e-beam, and the Ti:Sapphire enable the research and development of next generation of particle and photon sources.

\subsection{AWA}
A proposal to impliment a major upgrade of the Argonne Wakefield Accelerator (AWA) to the AWA-HE (HIGH ENERGY) Facility is under development . In several years, after completing the 500-MeV demonstrator, the AWA will become limited in its ability to continue advancing the SWFA Roadmap. Our vision is to continue to advance the Roadmaps by enacting a phased upgrade of the AWA facility so that by the end of the decade the SWFA method will be ready to present an SWFA CDR for a 3-TeV e+e- linear collider. Bunker expansion. Step 1 is to expand the AWA bunker length so that it occupies all available space in its current building. Energy upgrade. Step 2 is to add more RF stations and accelerating structures to increase the energy from 65 to 150 MeV. This will enable GV/m class acceleration gradients and a 3-GeV demonstrator.

\subsection{BELLA}
Upgrades are underway to the BELLA PW laser beamline (operating at 1 Hz) to allow the delivery of two synchronized pulses on target, enabling staging at the few GeV level, and a short focal length capability, enabling experiments at ultrahigh intensity (e.g., laser ion acceleration). The BELLA Center is also pursuing a new facility, kBELLA, consisting of a 1~kHz, few J, 30 fs, high average power laser for the demonstration of a high rep-rate, precision LPA and subsequent applications. This will function as a user facility with the kHz, GeV class electron beams and the intrinsically synchronized photon pulses being available for a wide variety of experiments in the basic and applied sciences. One possible candidate laser technology for kBELLA is coherent combining of fiber lasers, which holds promise for providing high average power at high efficiency. To this end, the BELLA Center has a very active R\&D program on coherent combining (in both time and space, as well as spectrally) of fiber lasers in collaboration with University of Michigan and LLNL. Furthermore, fiber lasers have the potential to provide high average power and high efficiency at 10 kHz and beyond, which is needed for a future LPA-based collider. The development of high-average power, high efficiency lasers is delineated as an important R\&D topic on the Advanced Accelerator Roadmap.

\subsection{FAST}
In the coming years, the IOTA ring will be upgraded with a proton injector and an electron lens/cooler~\cite{stancari-2021}. The research with protons and electrons will focus on nonlinear beam optics studies~\cite{antipov-2016}, investigation of space-charge effects~\cite{hwang-2020}, realization of Electron Lens with its broad research program~\cite{stancari-2021}, continuation of the OSC program~\cite{lebedev-2021}, and photon science opportunities~\cite{lobach-2021}. A significant required upgrade to the FAST facility is the addition of a laser system to enable research into electron-based coherent radiation generation and activities synergistic with advanced acceleration technologies.

\subsection{FACET-II}
As part of the current Snowmass process, the High Energy Physics community is evaluating a e-e+ collider options for precision measurements of the Higgs Boson. Advanced Accelerator technologies are being considered for the possibility of extending the energy reach of such a facility through high-gradient advanced accelerator-based energy boosters or afterburners. Additional studies are looking at the opportunities for Advanced Accelerator technologies to provide an order of magnitude increase in energy for discovery-oriented colliders in the future. In both of these scenarios it is critical to demonstrate a technique for accelerating high-intensity low-emittance positron bunches in plasma. The only experimental studies of positron acceleration in plasma have taken place at SLAC in the Final Focus Test Beam facility (FFTB) and FACET facilities. Novel ideas for positron acceleration using tailored beam drivers and/or plasma sources have been proposed using beam parameters envisioned for FACET-II. A plan is currently being developed to restore positron capabilities to FACET-II and, for the first time, provide the capability to deliver electrons and positrons to the experimental area simultaneously to enable studies of positron acceleration in electron beam driven wakefields.

\acknowledgments
We acknowledge support by the U.S. Department of Energy, Office of Science under contracts DE-AC02-06CH11357 with Argonne National Laboratory, DE-AC02-05CH11231 with Lawrence Berkeley National Laboratory, DE-AC02-07CH11359 with Fermi Research Alliance, DE-SC0018656 with Northern Illinois University, and DE-AC02-76SF00515 with SLAC National Accelerator Laboratory.

\bibliographystyle{unsrtnat}
\bibliography{biblio}
\end{document}